%% file: EEP_v3.tex
\documentclass[%
 reprint,
 superscriptaddress,
nobibnotes,
 amsmath,amssymb,
aps,
prl,
]{revtex4-1}

\usepackage{ulem}
\usepackage{graphicx}
\usepackage{txfonts}
\usepackage{siunitx}
\usepackage[hidelinks]{hyperref}

\usepackage{dcolumn}
\usepackage{bm}

\begin{document}
\title{Test of the Einstein Equivalence Principle near the Galactic Center Supermassive Black Hole}

\affiliation{
	Max Planck Institute for extraterrestrial Physics,
	Gie\ss enbachstra\ss e~1, 85748 Garching, Germany
}
\affiliation{
	LESIA, Observatoire de Paris, Universit\'e PSL, 
	CNRS, Sorbonne Universit\'e, Univ. Paris Diderot, 
	Sorbonne Paris Cit\'e, 5 place Jules Janssen, 92195 Meudon, France
}
\affiliation{
	Max Planck Institute for Astronomy, K\"onigstuhl 17, 
	69117 Heidelberg, Germany
}
\affiliation{
	$1^{\rm st}$ Institute of Physics, University of Cologne,
	Z\"ulpicher Straße 77, 50937 Cologne, Germany
}
\affiliation{
	Univ. Grenoble Alpes, CNRS, IPAG, 38000 Grenoble, France
}
\affiliation{
	Universidade de Lisboa - Faculdade de Ci\^encias, Campo Grande,
	1749-016 Lisboa, Portugal 
}
\affiliation{
	Faculdade de Engenharia, Universidade do Porto, rua Dr. Roberto
	Frias, 4200-465 Porto, Portugal 
}
\affiliation{
	European Southern Observatory, Casilla 19001, Santiago 19, Chile
}
\affiliation{
	Sterrewacht Leiden, Leiden University, Postbus 9513, 2300 RA
	Leiden, The Netherlands
}
\affiliation{
	Departments of Physics and Astronomy, Le Conte Hall, University
	of California, Berkeley, CA 94720, USA
}
\affiliation{
	School of Physics and Astronomy, Tel Aviv University, Tel Aviv
	69978, Israel
}
\affiliation{
	Max Planck Institute for Radio Astronomy, Auf dem H\"ugel 69, 53121
	Bonn, Germany
}
\affiliation{
	CENTRA - Centro de Astrof\'{\i}sica e
	Gravita\c c\~ao, IST, Universidade de Lisboa, 1049-001 Lisboa,
	Portugal
}
\affiliation{
	Center for Computational Astrophysics, Flatiron Instituete, 162 5th Ave.,
	New York, NY, 10010, USA
}

\author{A.~Amorim}
\affiliation{
	Universidade de Lisboa - Faculdade de Ci\^encias, Campo Grande,
	1749-016 Lisboa, Portugal 
}
\affiliation{
	CENTRA - Centro de Astrof\'{\i}sica e
	Gravita\c c\~ao, IST, Universidade de Lisboa, 1049-001 Lisboa,
	Portugal
}
\author{M.~Baub\"ock}
\affiliation{
	Max Planck Institute for extraterrestrial Physics,
	Gie\ss enbachstra\ss e~1, 85748 Garching, Germany
}
\author{J.P.~Berger}
\affiliation{
	Univ. Grenoble Alpes, CNRS, IPAG, 38000 Grenoble, France
}
\author{W.~Brandner}
\affiliation{
	Max Planck Institute for Astronomy, K\"onigstuhl 17, 
	69117 Heidelberg, Germany
}
\author{Y.~Cl\'{e}net}
\affiliation{
	LESIA, Observatoire de Paris, Universit\'e PSL, 
	CNRS, Sorbonne Universit\'e, Univ. Paris Diderot, 
	Sorbonne Paris Cit\'e, 5 place Jules Janssen, 92195 Meudon, France
}
\author{V.~Coud\'e~du~Foresto}
\affiliation{
	LESIA, Observatoire de Paris, Universit\'e PSL, 
	CNRS, Sorbonne Universit\'e, Univ. Paris Diderot, 
	Sorbonne Paris Cit\'e, 5 place Jules Janssen, 92195 Meudon, France
}
\author{P.T.~de~Zeeuw}
\affiliation{
	Sterrewacht Leiden, Leiden University, Postbus 9513, 2300 RA
	Leiden, The Netherlands
}
\affiliation{
	Max Planck Institute for extraterrestrial Physics,
	Gie\ss enbachstra\ss e~1, 85748 Garching, Germany
}
\author{J.~Dexter}
\affiliation{
	Max Planck Institute for extraterrestrial Physics,
	Gie\ss enbachstra\ss e~1, 85748 Garching, Germany
}
\author{G.~Duvert}
\affiliation{
	Univ. Grenoble Alpes, CNRS, IPAG, 38000 Grenoble, France
}
\author{M.~Ebert}
\affiliation{
	Max Planck Institute for Astronomy, K\"onigstuhl 17, 
	69117 Heidelberg, Germany
}
\author{A.~Eckart}
\affiliation{
	$1^{\rm st}$ Institute of Physics, University of Cologne,
	Z\"ulpicher Straße 77, 50937 Cologne, Germany
}
\affiliation{
	Max Planck Institute for Radio Astronomy, Auf dem H\"ugel 69, 53121
	Bonn, Germany
}
\author{F.~Eisenhauer}
\affiliation{
	Max Planck Institute for extraterrestrial Physics,
	Gie\ss enbachstra\ss e~1, 85748 Garching, Germany
}
\author{N.M.~F\"{o}rster~Schreiber}
\affiliation{
	Max Planck Institute for extraterrestrial Physics,
	Gie\ss enbachstra\ss e~1, 85748 Garching, Germany
}
\author{P.~Garcia}
\affiliation{
	Faculdade de Engenharia, Universidade do Porto, rua Dr. Roberto
	Frias, 4200-465 Porto, Portugal 
}
\affiliation{
	European Southern Observatory, Casilla 19001, Santiago 19, Chile
}
\affiliation{
	CENTRA - Centro de Astrof\'{\i}sica e
	Gravita\c c\~ao, IST, Universidade de Lisboa, 1049-001 Lisboa,
	Portugal
}
\author{F.~Gao}
\affiliation{
	Max Planck Institute for extraterrestrial Physics,
	Gie\ss enbachstra\ss e~1, 85748 Garching, Germany
}
\author{E.~Gendron}
\affiliation{
	LESIA, Observatoire de Paris, Universit\'e PSL, 
	CNRS, Sorbonne Universit\'e, Univ. Paris Diderot, 
	Sorbonne Paris Cit\'e, 5 place Jules Janssen, 92195 Meudon, France
}
\author{R.~Genzel}
\affiliation{
	Max Planck Institute for extraterrestrial Physics,
	Gie\ss enbachstra\ss e~1, 85748 Garching, Germany
}
\affiliation{
	Departments of Physics and Astronomy, Le Conte Hall, University
	of California, Berkeley, CA 94720, USA
}
\author{S.~Gillessen}
\affiliation{
	Max Planck Institute for extraterrestrial Physics,
	Gie\ss enbachstra\ss e~1, 85748 Garching, Germany
}
\author{M.~Habibi}
\thanks{Corresponding author: \href{mailto:mhabibi@mpe.mpg.de}{mhabibi@mpe.mpg.de}}
\affiliation{
	Max Planck Institute for extraterrestrial Physics,
	Gie\ss enbachstra\ss e~1, 85748 Garching, Germany
}

\author{X.~Haubois}
\affiliation{
	European Southern Observatory, Casilla 19001, Santiago 19, Chile
}
\author{Th.~Henning}
\affiliation{
	Max Planck Institute for Astronomy, K\"onigstuhl 17, 
	69117 Heidelberg, Germany
}
\author{S.~Hippler}
\affiliation{
	Max Planck Institute for Astronomy, K\"onigstuhl 17, 
	69117 Heidelberg, Germany
}
\author{M.~Horrobin}
\affiliation{
	$1^{\rm st}$ Institute of Physics, University of Cologne,
	Z\"ulpicher Straße 77, 50937 Cologne, Germany
}
\author{Z.~Hubert}
\affiliation{
	Univ. Grenoble Alpes, CNRS, IPAG, 38000 Grenoble, France
}
\author{A.~Jim\'enez~Rosales}
\affiliation{
	Max Planck Institute for extraterrestrial Physics,
	Gie\ss enbachstra\ss e~1, 85748 Garching, Germany
}
\author{L.~Jocou}
\affiliation{
	Univ. Grenoble Alpes, CNRS, IPAG, 38000 Grenoble, France
}
\author{P.~Kervella}
\affiliation{
	LESIA, Observatoire de Paris, Universit\'e PSL, 
	CNRS, Sorbonne Universit\'e, Univ. Paris Diderot, 
	Sorbonne Paris Cit\'e, 5 place Jules Janssen, 92195 Meudon, France
}
\author{S.~Lacour}
\affiliation{
	LESIA, Observatoire de Paris, Universit\'e PSL, 
	CNRS, Sorbonne Universit\'e, Univ. Paris Diderot, 
	Sorbonne Paris Cit\'e, 5 place Jules Janssen, 92195 Meudon, France
}
\affiliation{
	Max Planck Institute for extraterrestrial Physics,
	Gie\ss enbachstra\ss e~1, 85748 Garching, Germany
}
\author{V.~Lapeyr\`ere}
\affiliation{
	LESIA, Observatoire de Paris, Universit\'e PSL, 
	CNRS, Sorbonne Universit\'e, Univ. Paris Diderot, 
	Sorbonne Paris Cit\'e, 5 place Jules Janssen, 92195 Meudon, France
}
\author{J.-B.~Le~Bouquin}
\affiliation{
	Univ. Grenoble Alpes, CNRS, IPAG, 38000 Grenoble, France
}
\author{P.~L\'ena}
\affiliation{
	LESIA, Observatoire de Paris, Universit\'e PSL, 
	CNRS, Sorbonne Universit\'e, Univ. Paris Diderot, 
	Sorbonne Paris Cit\'e, 5 place Jules Janssen, 92195 Meudon, France
}
\author{T.~Ott}
\affiliation{
	Max Planck Institute for extraterrestrial Physics,
	Gie\ss enbachstra\ss e~1, 85748 Garching, Germany
}
\author{T.~Paumard}
\affiliation{
	LESIA, Observatoire de Paris, Universit\'e PSL, 
	CNRS, Sorbonne Universit\'e, Univ. Paris Diderot, 
	Sorbonne Paris Cit\'e, 5 place Jules Janssen, 92195 Meudon, France
}
\author{K.~Perraut}
\affiliation{
	Univ. Grenoble Alpes, CNRS, IPAG, 38000 Grenoble, France
}
\author{G.~Perrin}
\affiliation{
	LESIA, Observatoire de Paris, Universit\'e PSL, 
	CNRS, Sorbonne Universit\'e, Univ. Paris Diderot, 
	Sorbonne Paris Cit\'e, 5 place Jules Janssen, 92195 Meudon, France
}
\author{O.~Pfuhl}
\affiliation{
	Max Planck Institute for extraterrestrial Physics,
	Gie\ss enbachstra\ss e~1, 85748 Garching, Germany
}
\author{S.~Rabien}
\affiliation{
	Max Planck Institute for extraterrestrial Physics,
	Gie\ss enbachstra\ss e~1, 85748 Garching, Germany
}
\author{G.~Rodr\'iguez-Coira}
\affiliation{
	LESIA, Observatoire de Paris, Universit\'e PSL, 
	CNRS, Sorbonne Universit\'e, Univ. Paris Diderot, 
	Sorbonne Paris Cit\'e, 5 place Jules Janssen, 92195 Meudon, France
}
\author{G.~Rousset}
\affiliation{
	LESIA, Observatoire de Paris, Universit\'e PSL, 
	CNRS, Sorbonne Universit\'e, Univ. Paris Diderot, 
	Sorbonne Paris Cit\'e, 5 place Jules Janssen, 92195 Meudon, France
}
\author{S.~Scheithauer}
\affiliation{
	Max Planck Institute for Astronomy, K\"onigstuhl 17, 
	69117 Heidelberg, Germany
}
\author{A.~Sternberg}
\affiliation{
	School of Physics and Astronomy, Tel Aviv University, Tel Aviv
	69978, Israel
}
\affiliation{
	Center for Computational Astrophysics, Flatiron Instituete, 162 5th Ave.,
	New York, NY, 10010, USA
}
\author{O.~Straub}
\affiliation{
	LESIA, Observatoire de Paris, Universit\'e PSL, 
	CNRS, Sorbonne Universit\'e, Univ. Paris Diderot, 
	Sorbonne Paris Cit\'e, 5 place Jules Janssen, 92195 Meudon, France
}
\affiliation{
	Max Planck Institute for extraterrestrial Physics,
	Gie\ss enbachstra\ss e~1, 85748 Garching, Germany
}

\author{C.~Straubmeier}
\affiliation{
	$1^{\rm st}$ Institute of Physics, University of Cologne,
	Z\"ulpicher Straße 77, 50937 Cologne, Germany
}
\author{E.~Sturm}
\affiliation{
	Max Planck Institute for extraterrestrial Physics,
	Gie\ss enbachstra\ss e~1, 85748 Garching, Germany
}
\author{L.J.~Tacconi}
\affiliation{
	Max Planck Institute for extraterrestrial Physics,
	Gie\ss enbachstra\ss e~1, 85748 Garching, Germany
}
\author{F.~Vincent}
\affiliation{
	LESIA, Observatoire de Paris, Universit\'e PSL, 
	CNRS, Sorbonne Universit\'e, Univ. Paris Diderot, 
	Sorbonne Paris Cit\'e, 5 place Jules Janssen, 92195 Meudon, France
}
\author{S.~von~Fellenberg}
\affiliation{
	Max Planck Institute for extraterrestrial Physics,
	Gie\ss enbachstra\ss e~1, 85748 Garching, Germany
}
\author{I.~Waisberg}
\affiliation{
	Max Planck Institute for extraterrestrial Physics,
	Gie\ss enbachstra\ss e~1, 85748 Garching, Germany
}
\author{F.~Widmann}
\thanks{Corresponding author: \href{mailto:fwidmann@mpe.mpg.de}{fwidmann@mpe.mpg.de}}

\affiliation{
	Max Planck Institute for extraterrestrial Physics,
	Gie\ss enbachstra\ss e~1, 85748 Garching, Germany
}

\author{E.~Wieprecht}
\affiliation{
	Max Planck Institute for extraterrestrial Physics,
	Gie\ss enbachstra\ss e~1, 85748 Garching, Germany
}
\author{E.~Wiezorrek}
\affiliation{
	Max Planck Institute for extraterrestrial Physics,
	Gie\ss enbachstra\ss e~1, 85748 Garching, Germany
}
\author{S.~Yazici}
\affiliation{
	Max Planck Institute for extraterrestrial Physics,
	Gie\ss enbachstra\ss e~1, 85748 Garching, Germany
}
\affiliation{
	$1^{\rm st}$ Institute of Physics, University of Cologne,
	Z\"ulpicher Straße 77, 50937 Cologne, Germany
}

\collaboration{GRAVITY Collaboration}
\noaffiliation{}
\date{\today}%

\begin{abstract}
During its orbit around the four million solar mass black hole Sagittarius A* the star S2 experiences significant changes in gravitational potential. We use this change of potential to test one part of the Einstein equivalence principle: the local position invariance (LPI). We study the dependency of different atomic transitions on the gravitational potential to give an upper limit on violations of the LPI. This is done by separately measuring the redshift from hydrogen and helium absorption lines in the stellar spectrum during its closest approach to the black hole.
For this measurement we use radial velocity data from 2015 to 2018 and combine it with the gravitational potential at the position of S2, which is calculated from the precisely known orbit of S2 around the black hole. This results in a limit on a violation of the LPI of $|\beta_{He}-\beta_{H}| = (2.4 \pm 5.1) \cdot 10^{-2}$. The variation in potential that we probe with this measurement is six magnitudes larger than possible for measurements on Earth, and a factor ten larger than in experiments using white dwarfs. We are therefore testing the LPI in a regime where it has not been tested before.
\end{abstract}

\maketitle

\section{Introduction}
Since its publication in 1915 general relativity (GR) has been tested frequently and has so far passed all experimental tests \citep{Will2014}. Recently there has been an additional experiment in a new mass regime: For the first time it was possible to detect both the gravitational redshift and the transverse Doppler shift of a star moving on an elliptical orbit through the extreme gradient of the gravitational potential near a supermassive black hole \citep{GRAVITY2018}. This was possible by monitoring the orbit of the star S2 around the supermassive black hole Sagittarius A* (Sgr A*) over the last 26 years \citep[see e.g.][]{Ghez2008, Gillessen2009, Gillessen2017b}. So far all data taken for this experiment show excellent agreement with the predictions from GR. This work expands the previous tests of this experiment by testing the Einstein equivalence principle (EEP). The EEP states the universality of the coupling of gravity to matter and energy. Tests of the EEP are of great importance as many alternative theories of gravity and theories unifying gravity with other interactions predict violations of the EEP at high energies \citep{Damour1996, Flambaum2008}. The EEP consists of three main principles: the weak equivalence principle (WEP), the local position invariance (LPI), and the local Lorentz invariance \citep{Will1993, Will2014}.  From those three principles the local Lorentz invariance is best constrained, as no violations have been found down to $c_0^2/c^2-1 < 10^{-20}$ \citep{Chupp1989,Will2014}. It is therefore assumed to be valid for this work, while the the LPI is discussed in the following. The WEP or universality of free fall is not straight forward to test with our current approach \cite{Angelil2011}, which is discussed in more detail in the outlook.

\section{Galactic Center Experiment}\label{sec.gc}
Located at the very center of our galaxy is the bright radio source Sgr A*. The nuclear star cluster around it has been observed with high-resolution near-infrared (NIR) speckle and adaptive optics (AO) assisted imaging and spectroscopy over the past 26 years. This led to orbit determinations for $\approx$ 45 individual stars \citep{Schoedel2002,Schoedel2009,Ghez2003,Ghez2008,Eisenhauer2005,Gillessen2009,Gillessen2017b,Meyer2012,Boehle2016,Fritz2016}. These observations have demonstrated that the gravitational potential is dominated by a compact object at the center of the cluster. The mass of the object was measured by \cite{GRAVITY2018} to be (4.10 $\pm$ 0.03 ) x 10$^6$ M$_\odot$. 

The radio source Sgr A* is coincident with the center of mass to $<$ \SI{1}{\milli\arcsecond} \citep{Plewa2015}, and is itself very compact, with an upper limit on the radius of \SI{18}{\micro\arcsecond}, based on very long baseline interferometry at a wavelength of \SI{1.3}{\milli\meter} \citep{Krichbaum1998,Doeleman2008,Johnson2017}. In addition, Sgr A* shows, in comparison to extragalactic sources,  no intrinsic motion \citep{Reid2004,Reid2009}. This supports the interpretation that the compact radio source is coincident with the mass. Orbital motion of the centroids of the SgrA*'s near-infrared emission during bright 'flare states' suggest that the same mass inferred from the S2 orbit is also contained within \SIrange[range-units = single]{60}{90}{\micro\arcsecond} of the mean-position, or near the innermost stable orbit of a 4 million solar mass black hole \citep{Gravity2018b}. This all leads to the conclusion that Sgr A* is indeed a supermassive black hole \citep{Ghez2008,Genzel2010,Falcke2013}.

Of all the stars in the central cluster, the main-sequence B-star S2 is of special interest. With a near-infrared K-band magnitude of 14.2, S2 is one of the brightest stars in the innermost region around the black hole. It has an orbital period of 16.05 years and has its closest encounter with Sgr A* at a distance of 16.28 light hours or \SI{14.45}{\milli\arcsecond}. S2 also appears to be a single star \citep{Martins2008, Habibi2017, Chu2018}. The close encounter with Sgr A* and the comparatively short period make it the best available probe for post-Newtonian effects in the potential of the supermassive black hole \citep{GRAVITY2018}. One thing one might have to consider, is that S2 could come so close to the black hole that the star's properties change. However, the tidal disruption radius \citep{Hills1975} of the star S2, based on its stellar parameters \citep{Habibi2017}, is 100 times smaller than the star's periapsis distance. Therefore, we do not expect any strong tidal interactions between the star and the black hole. 

The GRAVITY Collaboration \citep{GRAVITY2018} showed that the data from S2 fulfills the predictions of general relativity when the gravitational redshift and the relativistic Doppler effect are taken into account. In Ref. \citep{GRAVITY2018} a scaling factor f for the first order parameterized post-Newtonian corrections (gravitational redshift and Doppler shift) is introduced, where f is zero for purely Newtonian physics and unity for GR. The measured f-factor of $f=0.90 \pm 0.09|_{\mbox{stat}} \pm 0.15|_{\mbox{sys}}$ is significantly inconsistent with pure Newtonian dynamics. The resulting f-value is getting more robust with more data added to the dataset. The same analysis as in \citep{GRAVITY2018}, but with additional data taken between June and September 2018, reduced the uncertainties in the f-value to $f=0.97 \pm 0.05|_{\mbox{stat}} \pm 0.05|_{\mbox{sys}}$ \citep{GRAVITY2019}.

\section{Local Position Invariance}\label{sec.lpi}
The main part of this work focuses on the LPI, which states that local nongravitational measurements are independent of their location in spacetime. To test this we use the star S2 as it moves on its eccentric orbit through the gravitational potential of Sgr A*. A violation of the LPI would imply a coupling of fundamental atomic constants, such as the fine structure constant, to the gravitational potential. LPI experiments can therefore be used to constrain coupling constants of different atomic properties \citep{Flambaum2007, Flambaum2008}. As such couplings are expected to be nonlinear it is especially important to perform such experiments with strong changes in potential.

According to the LPI, the gravitational redshift of a clock moving through a weak gravitational field ($\Phi/c^2 \ll 1$) with a varying potential $\Delta\Phi$, depends only on the change of the potential: $\Delta \nu / \nu = \Delta\Phi / c^2$, where $\nu$ is the clock frequency and $\Delta\nu$ the shift due to the gravitational potential. The formula implies that the shift in frequency does not depend on the internal structure of the clock, which is another way to formulate the LPI. To test this assumption one introduces a violation to the formula, commonly parametrized as $\beta$:
\begin{equation}
\frac{\Delta \nu}{\nu} = (1+\beta)~\frac{\Delta \Phi}{c^2}
\end{equation}
To test the LPI with a single type of clock one needs to compare two identical clocks in different gravitational potentials. Alternatively one can measure the frequency change of two non-identical clocks moving through a time-dependent potential $\Phi(t) = \Phi_0 + \Delta\Phi(t)$. In this case a violation of the LPI would again be visible in the fractional frequency difference:
\begin{equation}
\Delta \left(\frac{\Delta \nu}{\nu}\right) = \frac{\Delta \nu_2}{\nu_2} - \frac{\Delta \nu_1}{\nu_1} = (\beta_2-\beta_1)~\frac{\Delta \Phi(t)}{c^2} = \Delta\beta ~\frac{\Delta \Phi}{c^2}
\label{equ.lpi}
\end{equation}
By measuring the frequency change of two clocks moving through a potential one can therefore constrain $\Delta \beta$. Such \textit{null redshift experiments} are regularly done on Earth using the gravitational potential of the Sun, which varies over the timescale of a year, due to Earth's eccentric orbit \citep[see e.g.][]{Ashby2007, Agachev2011, Peil2013, Dzuba2017}. The annual potential variation due to this eccentric motion is $\Delta\Phi / c^2 = 3.3 \cdot 10^{-10}$. The most stringent limit on a violation of the LPI so far is given by Ref. \citep{Peil2013}, from a comparison of hydrogen masers with rubidium clocks. From this measurement a value of $|\beta_H - \beta_{Rb}| = (2.7 \pm 4.9) \cdot 10^{-7}$ is measured. To get to such a low limit it is necessary to measure the frequency change of atomic transitions with a precision on the order of $\Delta \nu/\nu \approx 10^{-17}$. The most stringent astronomical tests of the LPI were done by a comparison of measured wavelength shift in white dwarf spectra directly to laboratory wavelengths, to get a constraint on variations of the fine structure constant \cite{Berengut2013,Ong2013,Bainnbridge2017}. In the experiments with white dwarfs a potential difference of approximately $10^{-5}$ is reached, which is much higher than that possible for earthbound experiments. However, it is still roughly an order of magnitude lower than the potential difference observed for S2 orbiting around Sgr A*.

\subsection{Measurement}
The data for the Galactic center experiment were mainly taken with the European Southern Observatory's Very Large Telescope and Very Large Telescope Interferometer, using the three instruments NACO \citep{Lenzen1998,Rousset1998}, SINFONI \citep{Eisenhauer2003, Bonnet2004}, and GRAVITY \citep{GRAVITY2017}. The NACO images provided the time-dependent 2D projected positions of the stars in the nuclear star cluster. Those positions are then calibrated relative to the radio frame of the Galactic center \citep{Reid2007}. The unique astrometric precision of $\sim$ \SI{50}{\micro\arcsecond} obtained with GRAVITY directly adds the 2D projected separation of S2 and Sgr A* to the data set. SINFONI then adds spectroscopic measurements of the stars in order to measure their line-of-sight velocity \citep[for more details on the data and the data analysis see Ref.][]{GRAVITY2018}. The combination of the data is then used to fit the full orbit of S2 around the central black hole \citep{Gillessen2009,GRAVITY2018}. For this work we use the S2 orbit \citep{GRAVITY2018} to calculate the gravitational potential at the position of S2. This is done by calculating the Newtonian potential for the separation $d(t)$ between S2 and Sgr A*: $\Phi(t) = GM/d(t)$, with M being the mass of the black hole. For this calculation we can neglect all other stars in the area, as their masses are negligible in comparison to Sgr A*. Furthermore we can use a Newtonian description for the potential, as the first relativistic correction term would be from the Schwarzschild metric, which is so small that it is not yet relevant for the orbit fit \cite{Alexander2005, Gillessen2009}. In the three years leading up to the pericenter passage of S2 around the super massive black hole Sgr A*, the gravitational potential experienced by the star changes by $\Delta\Phi / c^2 = 3.2 \cdot 10^{-4}$.

\begin{figure}
\centering
\includegraphics[width=\columnwidth]{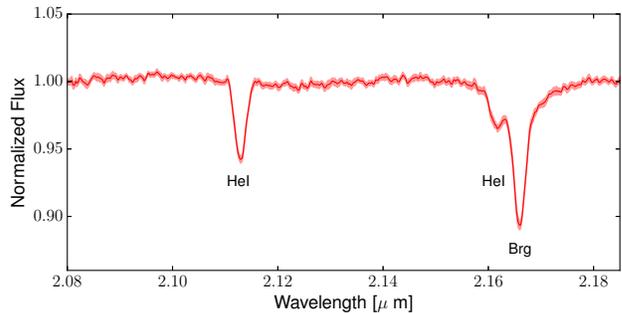}
\caption{High signal-to-noise spectrum of the star S2 in the astronomical K-Band. The spectrum has been produced by combining data from 12 years of observations \citep[adapted from][]{Habibi2017}.}
\label{fig.S2deep}
\end{figure}

In addition to the gravitational potential \cite{GRAVITY2018} the data used for this work are the K-Band (\SIrange[range-units = single]{2.0}{2.5}{\micro\meter}) spectra of S2 obtained with SINFONI. These spectra are used to measure the line-of-sight velocity of S2. In the K-band S2 has two dominant absorption features: The strongest line is the Br$\gamma$ line (hydrogen transition n = 7 - 4) with a vacuum wavelength of \SI{2.1661}{\micro\meter}. The second feature is the helium line around \SI{2.1125}{\micro\meter}. This line is not a single feature but a blend of the He I triplet at \SI{2.1120}{\micro\meter} (3p $^3$P$^0$ – 4s $^3$S) and the He I singlet at \SI{2.1132}{\micro\meter} (3p $^1$P$^0$ – 4s $^1$S). The weighting of the two features depends on the atmospheric parameters and the rotational velocity of the star \citep{Habibi2017}. In an individual spectrum at our resolution they appear as a single feature. In a typical observation of 1 hour the helium and hydrogen feature can be detected at $>$ 5$\sigma$.  A combined spectrum with a high signal-to-noise ratio (SNR) from Ref. \cite{Habibi2017} is shown in \autoref{fig.S2deep}. On the left shoulder of the hydrogen line is another helium line at \SI{2.161}{\micro\meter}, which is much weaker than the hydrogen line (flux ratio of 1 to 4 in the high SNR spectrum). In an individual dataset this line is just above the noise level. It is therefore not a dominant feature and does not influence the velocity measurement from the hydrogen line.

\begin{figure}
\centering
\includegraphics[width=\columnwidth]{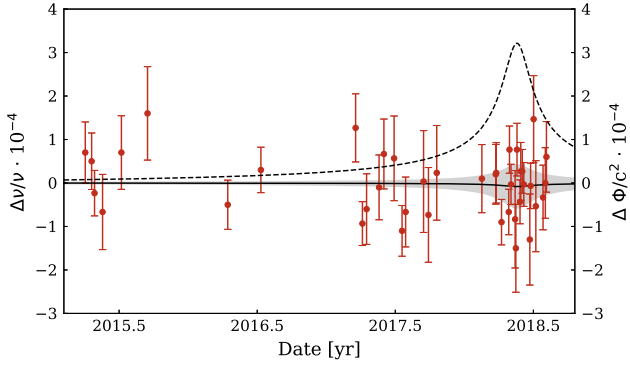}
\caption{Difference in frequency change for the helium and the hydrogen line as red dots. The dimensionless gravitational potential is shown as a dashed black line. The solid black line shows $\Delta \beta \cdot \Delta\Phi / c^2$, where $\Delta \beta$ is fitted to the data. The gray area shows the 3 sigma values from the fit.}
\label{fig.fit}
\end{figure}

After extracting the spectrum of S2 from the SINFONI data, we usually measure the star's velocity with a combination of a fit to the Br$\gamma$ line and a cross correlation of the whole K-band with the high SNR spectrum shown in \autoref{fig.S2deep} \citep[for more details see Ref.][]{GRAVITY2018}.  For this work we use a slightly different approach. We divide the spectrum into two parts, one containing the He feature and the other one the Br$\gamma$ line. Both parts are individually cross correlated with their corresponding part of the high SNR spectrum. By doing this we get two velocities for each spectrum: one from the helium line and one from the hydrogen line. In other words, we have a helium and a hydrogen clock moving through the varying gravitational potential during the pericenter passage of S2. By measuring the difference in frequency change for both clocks we are able to give an upper limit on the LPI violation during the pericenter passage. The values for the velocity difference $(\mbox{v}_{He} - \mbox{v}_{H})/c = \Delta \nu_{He}/\nu_{He} - \Delta \nu_{H}/\nu_{H}$ are shown in \autoref{fig.fit}, together with the gravitational potential at the position of S2.

The uncertainty of the datapoints in \autoref{fig.fit} is calculated from several contributions. The first is the calibration error of the wavelength. During the data reduction the wavelength calibration of each individual data frame is fine-tuned by a set of OH lines in the K-Band. The scatter of the line position from their expected velocity after the fine tuning is below \SI{5}{\kilo\meter\per\second}, which is then used as the uncertainty for the measured wavelength. This is calculated for each spectrum individually by fitting the atmospheric OH lines. A second contribution is the uncertainty of the cross correlation, determined from the uncertainty of the cross-correlation peak position. A third error originates from the extraction of the spectrum. As SINFONI is an integral field spectrograph the final result of the data reduction is a 3D cube, where two dimensions are the image axes and the third is the spectrum for each pixel. To get a spectrum of a star one has to select the source and background pixels in the image plane. This is the source of a third uncertainty as different masks can lead to slightly different results in the velocity. We account for this by calculating the velocity from different reasonable masks and use the scatter in the result as an estimate of uncertainty. The uncertainty of one velocity measurement is then the quadratic sum of these three contributions. This is done for Br$\gamma$ and He I individually. The final value used in this analysis is then the difference of the two velocities with the quadratic sum of the uncertainties. This might slightly overestimate the error as the calibration error should be the same for both measurements, but is accounted for twice. However, this does not have a big influence as it is the least dominant error source.

To get an upper limit on the LPI violation we use \autoref{equ.lpi} to fit the potential to the data points shown in \autoref{fig.fit}. In the fit $\Delta\beta = \beta_{He}-\beta_{H}$ is left as a free parameter. The fitted value of $\Delta\beta$ is:
\begin{equation}
\Delta\beta = |\beta_{He}-\beta_{H}| = (2.4 \pm 5.1) \cdot 10^{-2}
\end{equation}
Where the given error is the 1 $\sigma$ confidence interval of the fit. We can place an upper limit on the violation of the LPI in the strong gravitational field of the supermassive black hole of $\Delta\beta \leq 5 \cdot 10^{-2}$. The result is consistent with $\Delta\beta$ = 0. The fit is shown together with the data in \autoref{fig.fit}. The $\chi^2$ analysis of the fit shows a reduced $\chi^2$ of 0.91. In comparison, $\beta$ = 0 results in a $\chi^2_{red}$ of 0.89. Under the assumption that the $\chi^2$ distribution is approximately Gaussian it has a variance of $\sigma = \sqrt{2/N} = 0.22$. Therefore both values for $\chi^2_{red}$ lie within the one sigma range of $\chi^2_{red} = 1$ and the $\chi^2_{red}$ values cannot be used to distinguish between the models.

\begin{figure}
\centering
\includegraphics[width=\columnwidth]{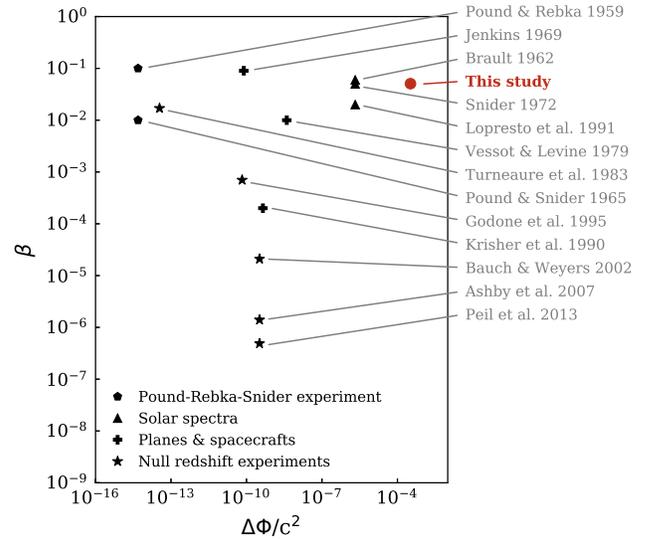}
\caption{Comparison of selected tests of the LPI with gravitational redshift. Plotted is the variation in potential, which is tested against the measured limit on a violation. The different symbols mark the Pound-Rebka-Snider experiments \citep{Pound1959, Pound1965}, tests from solar spectral lines \citep{Brault1962, Snider1972, Lopresto1991}, tests on rockets and spacecrafts \citep{Jenkins1969,Vessot1979,Krisher1990}, and null redshift experiments \citep{Turneaure1983, Godone1995, Bauch2002,Ashby2007,Peil2013}}
\label{fig.comparison}
\end{figure}

While our result is not competitive with current experiments on earth, the change in gravitational potential experienced by S2 on its orbit from early 2015 to its pericenter passage in May 2018 is $\Delta\Phi/c^2 = 3.2 \cdot 10^{-4}$. This is a regime which has not been reached by any other experiment and we therefore test the LPI at a potential difference which has not been tested before this work (see \autoref{fig.comparison}) \citep{Will2014}.

As mentioned in the introduction, a violation of the LPI would imply a coupling of fundamental atomic constants to the gravitational potential. Atomic clock measurements are therefore used to constrain coupling constants of different atomic properties \citep{Flambaum2008}. This can for example be done  for the coupling of the fine structure constant $\alpha$ \citep{Dzuba2017} or for the electron-to-proton mass ratio m$_e$/m$_p$ and the ratio of the light quark mass to the quantum chromodynamics length scale \cite{Peil2013}. In principle one could also use our measurement of $\beta$ to constrain these coupling constants. However, a single measurement of $\Delta\beta$ is not sufficient for that. One can overcome this by combining different measurements from different atomic species \citep{Peil2013}, or by using computational techniques to calculate the relativistic perturbation of the energy levels for the observed transitions \citep{Dzuba2017}. In the present case, the S2 helium absorption line is a doublet and the transitions are not isolated enough that a specific model of the transition would yield further information. We therefore cannot make any further statements than the pure limit on the violation of the LPI.

\section{Outlook}
This measurement demonstrates that the data from stars orbiting a black hole can be used for testing the LPI. Looking forward this also opens possibilities for the next pericenter passage of S2 in 2034. At that point the Extremely Large Telescope (ELT) will be fully operational. With a telescope diameter of more than four times the one from the VLT, the ELT will collect more than twenty times more light. The first light instrument MICADO \citep{Davies2016} will include a slit spectrograph with a resolving power of R $\geq$ 10000. This is more than six times higher than what we currently achieve with SINFONI (R = 1500 in the used mode). One can therefore use the ELT to measure S2's spectrum with higher resolution and with higher SNR. This would allow a velocity measurement of S2 in the H-Band, which currently has a too low SNR for velocity measurements from individual data frames. In the H-Band there is a narrow helium line (He I at \SI{1.7002}{\micro\meter}) as well as a series of hydrogen lines \citep{Habibi2017}, which can be used to significantly improve the velocity measurement. Unlike hydrogen, the He lines are not sensitive to the stellar pressure broadening, providing sharper atomic lines to measure the velocity with high accuracies.

With the high sensitivity of the ELT it is also possible to make the same measurement for fainter late type (K \& M type) stars. The infrared spectrum of these stars shows several sharp metal lines, including different isotopes, as well as series of rotational–vibrational bands of CO molecule \citep{Wallace1997}.  With a high resolution spectrograph such as the planned HIRES \citep{Marconi2016}, with a resolving power of R = 130000 and a very high calibration accuracy a velocity measurement of the order of \si{\meter\per\second} would be possible. This would allow a measurement of $\frac{\Delta \nu}{\nu}$ in the order of $10^{-8}$. For a star on a similar orbit as S2 this would translate in a factor of 10$^4$ more restrictive limit on the LPI and velocities from different atoms could be used to directly constrain coupling parameters. Interesting stars for this are for example S21 or S38 which are both in a comparably short orbit around SgrA* \citep[37 and 19 years, see][]{Gillessen2017b}, or even fainter stars in closer orbits which might be discovered with the ELT.

This would also open the possibility to test the third part of the EEP, the WEP, also known as universality of free fall. It states that inertial and gravitational mass are equivalent. In principle, one can use a gravitational redshift experiment to test the WEP, under the assumption that special relativity is fully valid \cite{Schiff1960}. However, in order to do so one has to precisely know the gravitational field, as otherwise a violation could be absorbed as a constant factor in the gravitational potential. A solution for this could be to use different stars around Sgr A*. In this case one star can be used to test the WEP and the others to measure the mass of Sgr A* separately \cite{Angelil2011}. At the moment this would be a rather imprecise measurement, as the current best mass measurement of Sgr A* is from S2 itself. This is a problem which is very likely to be solved with future observations and facilities. One solution would be the discovery of a star in closer orbit around SgrA*, either with GRAVITY \cite{Waisberg2018} or the ELT. The combination of S2 and a closer star can then be used to measure the mass of SgrA* and test the WEP individually. However, even without a star on a very close orbit, the ELT will allow more precise measurements of the already observed orbits of S-stars. With better orbit measurement of other close S-stars, such as S38, one can then test the WEP.

\section{Conclusion}
With this paper we continued the analysis of the data presented by the GRAVITY Collaboration \citep{GRAVITY2018} to give constraints on the LPI. We used the helium and hydrogen transitions in the spectrum of S2 as individual clocks, to give a constraint on a violation of the LPI. The results are consistent with the LPI and give an upper limit to a violation  of $5\cdot 10^{-2}$. This limit is in absolute numbers less stringent than the current most precise tests \citep{Peil2013}. Our experiment however tests the LPI close to a central black hole with 4 million solar masses, in a potential which is 10$^6$ times larger than accessible to terrestrial experiments. It is currently the most extreme test of the LPI and is fully consistent with it.

\section{Acknowledgments}
GRAVITY is developed in a collaboration by the Max Planck Institute for extraterrestrial Physics, LESIA of Paris Observatory/CNRS/Sorbonne Universit\'e/University Paris Diderot and IPAG of Universit́\'e Grenoble Alpes/CNRS, the Max Planck Institute for Astronomy, the University of Cologne, the CENTRA—Centro de Astrofisica e Gravita\c{c}\~{a}o, and the European Southern Observatory. We are very grateful to our funding agencies (MPG, ERC,CNRS, DFG, BMBF, Paris Observatory, Observatoire des Sciences de l’Universde Grenoble, and the Funda\c{c}\~{a}o para a Ci\^{e}ncia e Tecnologia), to ESO and the ESO/Paranal staff,and to the many scientific and technical staffmembers in our institutions who helped to make NACO, SINFONI, and GRAVITY a reality. S.G. acknowledges support from ERC starting grant No. 306311. F. E. and O. P. acknowledge support from ERC synergy Grant No. 610058 (Black-HoleCam). J. D., M. B., and A. J.-R. were supported by a Sofja  Kovalevskaja  award  from  the  Alexander  von Humboldt foundation. A. A. \& P. G. acknowledge support from FCT-Portugal with reference UID/FIS/00099/2013.

\input{EEP_v2.bbl}

\end{document}

%% file: EEP_v2.bbl
%